\documentclass[lettersize,journal]{IEEEtran}
\usepackage[utf8]{inputenc}
\usepackage{amsmath,amsfonts}
\usepackage{algorithmic}
\usepackage{algorithm}
\usepackage{array}
\usepackage[caption=false,font=normalsize,labelfont=sf,textfont=sf]{subfig}
\usepackage{textcomp}
\usepackage{stfloats}
\usepackage{url}
\usepackage{verbatim}
\usepackage{graphicx}
\usepackage{cite}
\usepackage{xcolor}
\usepackage{glossaries}
\title{Taming Aerial Communication with\\Flight-assisted Smart Surfaces in 6G Era}

\author{Francesco Devoti,~\IEEEmembership{Member,~IEEE,} Placido Mursia,~\IEEEmembership{Member,~IEEE,} \\ Vincenzo Sciancalepore,~\IEEEmembership{Senior Member,~IEEE,} Xavier Costa-Pérez,~\IEEEmembership{Senior Member,~IEEE}
\thanks{F. Devoti, P. Mursia, and V. Sciancalepore are with NEC Laboratories Europe GmbH, 69115 Heidelberg, Germany (e-mails: \{francesco.devoti, placido.mursia, vincenzo.sciancalepore\}@neclab.eu).}
\thanks{X. Costa-Pérez is with NEC Laboratories Europe GmbH, Heidelberg, Germany, and i2CAT Foundation and ICREA, Barcelona, Spain. (e-mail: xavier.costa@ieee.org).}
\thanks{This work was supported by the European Union H-2020 Project RISE-6G under grant 101017011.}
}

\newacronym{3d}{$3$D}{three-dimensional}
\newacronym{ai}{AI}{artificial intelligence}
\newacronym{ap}{AP}{access point}
\newacronym{bs}{BS}{base station}
\newacronym{ber}{BER}{bit error rate}
\newacronym{csi}{CSI}{channel state information}
\newacronym{iot}{IoT}{internet of things}
\newacronym{los}{LoS}{line-of-sight}
\newacronym{mec}{MEC}{multi-access edge computing}
\newacronym{mimo}{MIMO}{multiple input, multiple output}
\newacronym{ml}{ML}{machine learning}
\newacronym{sca}{SCA}{successive convex approximation}
\newacronym{sdr}{SDR}{semidefinite relaxation}
\newacronym{mmw}{mmWave}{millimeter-wave}
\newacronym{thz}{THz}{terahertz}
\newacronym{noma}{NOMA}{non-orthogonal multiple access}
\newacronym{ris}{RIS}{reconfigurable intelligent surface}
\newacronym{snr}{SNR}{signal-to-noise ratio}
\newacronym{uav}{UAV}{unmanned aerial vehicle}
\newacronym{hap}{HAP}{high-altitude platform}
\newacronym{fwa}{FWA}{fixed-wing aircraft}
\newacronym{ue}{UE}{user equipment}
\newacronym{urllc}{URLLC}{ultra-reliable low-latency communication}
\newacronym{kpi}{KPI}{key performance indicator}
\newacronym{5g}{5G}{fifth generation}
\newacronym{a2g}{A2G}{air-to-ground}
\newacronym{soa}{SoA}{state of the art}
\newacronym{qos}{QoS}{quality of service}
\newacronym{aoa}{AoA}{angle of arrival}
\newacronym{aod}{AoD}{angle of departure}
\newacronym{tdma}{TDMA}{time division multiple access}
\newacronym{gps}{GPS}{global positioning system}

\begin{document}

\setlength{\textfloatsep}{3pt}
\maketitle

\begin{abstract}
Aerial communication is gradually taking an assertive role within common societal behaviors by means of unmanned aerial vehicles (UAVs), high-altitude platforms (HAPs), and fixed-wing aircrafts (FWAs). Such devices can assist general operations in a diverse set of heterogeneous applications, such as video-surveillance, remote delivery and connectivity provisioning in crowded events and emergency scenarios.
Given their increasingly higher technology penetration rate, telco operators started looking at the sky as a new potential direction to enable a three-dimensional (3D) communication paradigm. 

However, designing \emph{flying mobile stations} involves addressing a daunting number of challenges, such as an excessive on-board control overhead, variable battery drain and advanced antenna design. To this end, the newly-born \emph{Smart Surfaces} technology may come to help: \emph{reconfigurable intelligent surfaces (RIS)} may be flexibly installed on-board to control the terrestrial propagation environment from an elevated viewpoint by involving low-complex and battery-limited solutions. In this paper, we shed light on novel RIS-based use-cases, corresponding requirements, and potential solutions that might be adopted in future aerial communication infrastructures. 
\end{abstract}

\section{Introduction}

\glsresetall 



Aerial communications, whereby flying devices such as \glspl{uav}, \glspl{fwa} or \glspl{hap} are used to facilitate the communication between two given nodes in the network, are a well-established technology capable of providing powerful yet highly flexible platforms to enhance and complement current \gls{5g} and beyond network infrastructures (6G)~\cite{Moza19}. {Such kind of deployments, denoted as \emph{\gls{a2g}} networks, are envisioned as an integral part of future \gls{iot} networks thanks to their agility to act as portable on-demand \glspl{bs}, flexibility to move in space, and an overall increased probability of high end-to-end channel quality. As a matter of fact, unlike fixed \glspl{bs}, flying devices can be exploited on-demand and reused for different applications, e.g., assisting rescue operations after natural disasters via user localization and (re)establishing a damaged or unavailable network infrastructure~\cite{albanese2021responders}. In addition, they are able to hover avoiding obstacles that may cause blockage. Thus, significantly increasing the probability of establishing a \gls{los} with \textit{terrestrial} users, giving rise to the new \gls{3d} communication paradigm~\cite{Calvanese2020} in the 6G era.}

However, despite its potentials and envisioned business opportunities, aerial communications are hindered by the limited power budget {and maximum carrying load} of flying devices. E.g., in the case of \glspl{uav} it poses significant constraints on its feasibility: carrying bulky, heavy, and power-hungry equipment such as active antennas on board results in high design complexity and capital expenditure; further limiting the application range of aerial communications. {Moreover, conventional \gls{a2g} networks require signal processing capabilities on-board, e.g., for channel estimation or precoding, and frequent backhaul communications, which further compound the design complexity of such systems.}
\begin{figure}
    \centering
    \includegraphics[width=\linewidth]{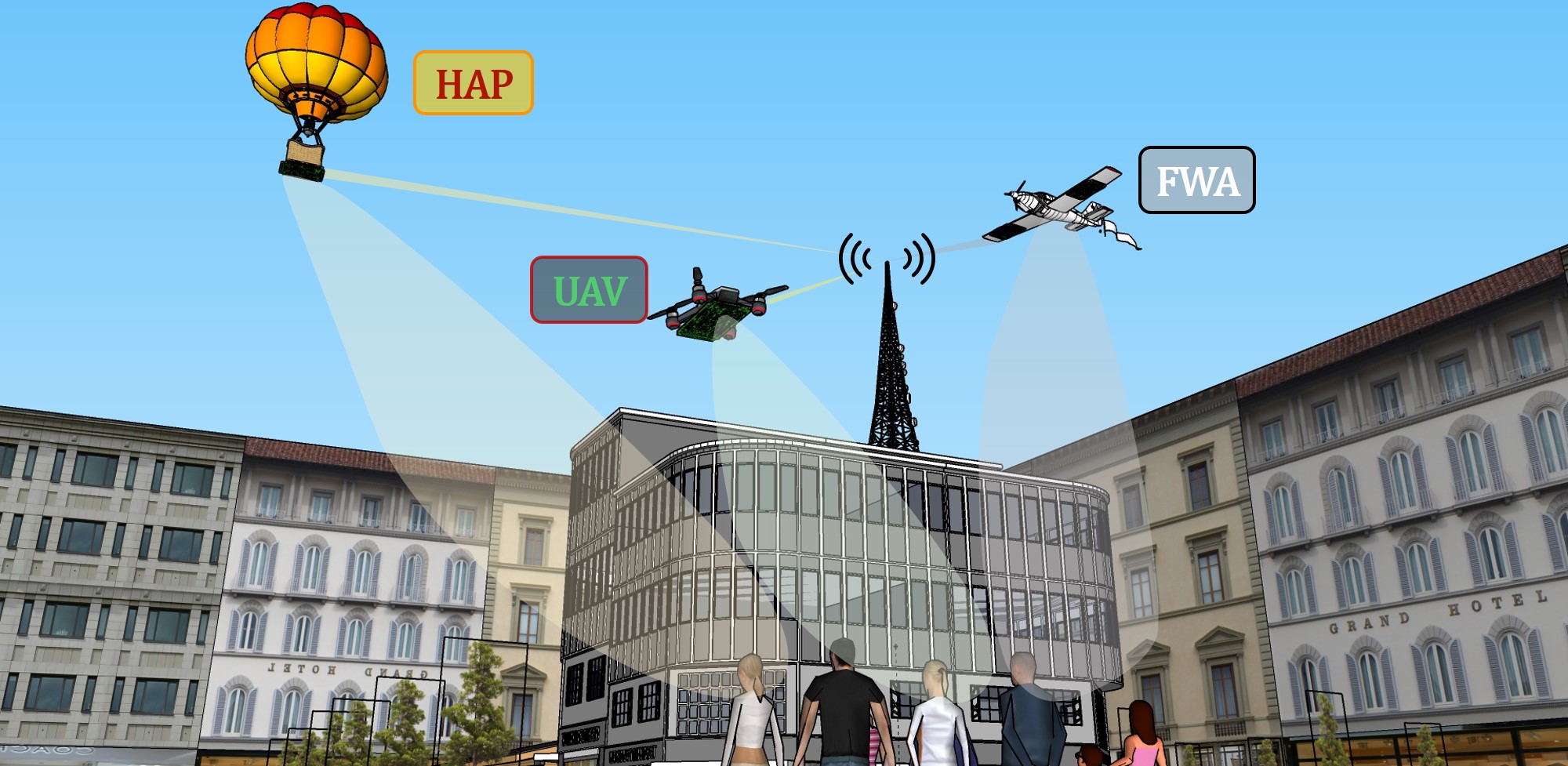}
    \caption{\acrshort{ris}-based \acrshort{3d} communication environment with \acrshort{uav}s, \acrshort{hap}s and \acrshort{fwa}s.}
    \label{fig:scenario_intro}
\end{figure}

To overcome the above-mentioned issues, there has been growing interest in the emerging \gls{ris} technology, which is widely recognized as the means to greatly improve the quality of wireless communication links towards 6G networks (\gls{a2g} communications) as depicted in Fig.~\ref{fig:scenario_intro}~\cite{Alf20}. \glspl{ris} are two-dimensional surfaces divided into unit cells spaced at a sub-wavelength distance, which can reflect the incoming signal by adding a given tunable phase shift. As a result, the reflected signal components can be combined constructively at the intended receiver position and destructively elsewhere {in a nearly-passive way~\cite{rise_commag21}}.
Moreover, \glspl{ris} can be fabricated to be extremely lightweight {(hundreds of grams) and consume as little as tens of milliwatts, as opposed to conventional \glspl{bs} that weight few Kilograms and consume tens of Watts. Thereby, making them suitable for being rapidly deployed in B5G/6G networks at affordable costs.}

{While the opportunities of \gls{ris}-aided \gls{a2g} networks are well-understood, several design aspects still need to be addressed in order to unlock its full potential.
Indeed, the joint optimization of the \gls{ris} and \gls{bs} configuration, as well as the aerial device trajectory, require an adaptive level of control and signaling able to cope with the degree of mobility of the specific application scenario, including: unpredictable movements typical of aerial objects, corresponding maneuvering and a limited power budget available on board. All such aspects pose strict limitations on the \gls{a2g} control architecture, making its design extremely challenging.
}

{In this paper, we identify the main optimization aspects to be considered when designing a \gls{ris}-aided \gls{a2g} network by describing its main characteristics and corresponding \gls{soa} solutions. We further shed light on technical challenges and open research directions and characterize two relevant operating scenarios.}
We finally deliver the message that \emph{while joint exploitation of \gls{ris} energy-efficient beamforming capabilities and flexible and agile aerial communications requires a careful design of the control architecture, it might significantly boost the overall communication performance}.

\section{RIS-Aided A2G Networks}


\Glspl{ris} can be integrated in the context of aerial communications in two different ways, namely \textit{i)} \textit{terrestrial} \gls{ris}, i.e., by mounting them on the facade of buildings in order to assist in the communication to/from the flying object and \textit{ii)} \textit{aerial} \gls{ris}, i.e., by employing them as substitutes to bulky active components such as conventional \glspl{bs} on board the flying device~\cite{Pang21}. However, while deploying terrestrial \glspl{ris} can help alleviate the total power consumption, it does not fundamentally solve the problem of limited operation range of e.g. \glspl{uav}. On the other side, by allowing the \gls{uav} to carry an inherently light-weight and passive device on board such as an \gls{ris}, nearly all the available power can be devoted to enlarge the range of operation while simultaneously achieving highly selective beamforming. The \gls{ris} itself might even be (partially) powered by harvesting power from the incoming signals, which further demonstrates the feasibility of aerial \glspl{ris}~\cite{Zha19}. Moreover, terrestrial \glspl{ris} can only achieve $180^\circ$ reflection angles, compared to the full-angle $360^\circ$ of aerial \glspl{ris}, which allows to effectively cover the intended two-dimensional target area. Therefore, we focus on aerial \glspl{ris} as depicted in Fig.~\ref{fig:scenario_intro} and describe the main characteristics, involved opportunities, and corresponding challenges.


The link budget of an \gls{a2g} network is directly proportional to the signal wavelength and the number of \gls{ris} elements, and inversely proportional to the distance to the user~\cite{Alf21}. On the one hand, the increasing demand for higher communication rates and the need to support a massive number of users is pushing future wireless networks towards higher frequency bands, such as \gls{mmw} or \gls{thz}. Hence, the signal wavelength is typically small and will likely tend to diminish, which further compounds the problem of correctly designing \gls{a2g} networks. On the other hand, while the received power at the ground user can be linearly increased by making \gls{ris} elements dense, \glspl{uav} can carry only a limited payload thus resulting in a maximum feasible \gls{ris} dimension. Note that to overcome this problem, \gls{ris}-equipped \gls{uav} \textit{swarms} can be deployed, i.e., a group of aerial \glspl{ris} that cooperate together~\cite{Sha21}. However, such cooperation comes at a significantly increased cost in terms of control overhead and complexity. It is thus of paramount importance to optimize the location of the aerial platform in order to obtain the best possible propagation conditions to the ground users in terms of increased \gls{los} probability and limited pathloss.


Beamforming at the \gls{ris} can increase the received power at the intended location up to a maximum theoretical factor equal to the square of the number of elements. In order to take full advantage of such tremendous capabilities, advanced \gls{3d} \textit{passive} beamforming algorithms and techniques at the \gls{ris} are needed, especially in the multi-user scenario, which requires addressing complex non-convex problems.

Moreover, current \gls{5g} \glspl{bs} are usually equipped with a multi-antenna array, which endows them of \textit{active} beamforming capabilities. The optimization of both active and passive beamforming vectors is intrinsically coupled with the optimization of the position of the flying device and its evolution over time (i.e., its trajectory). 
Such joint optimization is in general intractable and requires to resort to suboptimal approaches or heuristics. In particular, an efficient approach is decoupling the two problems in an alternating fashion. The trajectory optimization is efficiently solved via \gls{ml} tools such as deep networks, while the passive beamforming optimization problem can be tackled via \gls{sca} or \gls{sdr}~\cite{Kha21}. In the multi-user scenario, one can alternatively schedule users with a \gls{tdma} scheme~\cite{LiY21} or, in order to improve fairness and robustness, consider \textit{max-min} approaches thereby assigning at each user a given figure of merit such as, e.g., the \gls{snr} or the \gls{ber}, and aiming at maximizing the worst figure among the target users~\cite{Lu20_2}.


The core of the joint trajectory and beamforming optimization in \gls{a2g} networks is linked to the acquisition and tracking of \gls{csi}. Indeed, an accurate system optimization requires the estimation of both the channel from the \gls{bs} to the flying device and the channel from the latter to the target user. However, \glspl{ris} are passive structures with no estimation nor processing capabilities, which poses several constraints on the applicability of classical pilot-based channel estimation techniques~\cite{Hu21}. In addition, in real-life scenarios the users have a certain degree of mobility, which presents the highly challenging problem of tracking the evolution of the channel vectors over time. In this regard, it is essential to have accurate channel modeling that depends on few key system parameters. Thanks to the high \gls{los} probability associated with the altitude of the aerial platform thereby enabling the \gls{3d} communication paradigm~\cite{Calvanese2020}, the channel vectors are completely characterized by the distance travelled by the signal and the \gls{aoa}/\gls{aod} at the various entities in the network~\cite{Jian20}. It is thus possible to realize efficient \gls{csi} acquisition in \gls{a2g} networks by tracking the location of both the users and the flying device, which is feasible with a sufficient degree of control signalling. 

\section{A2G Control Architecture}

\begin{figure*}
    \centering
    \includegraphics[width=\linewidth]{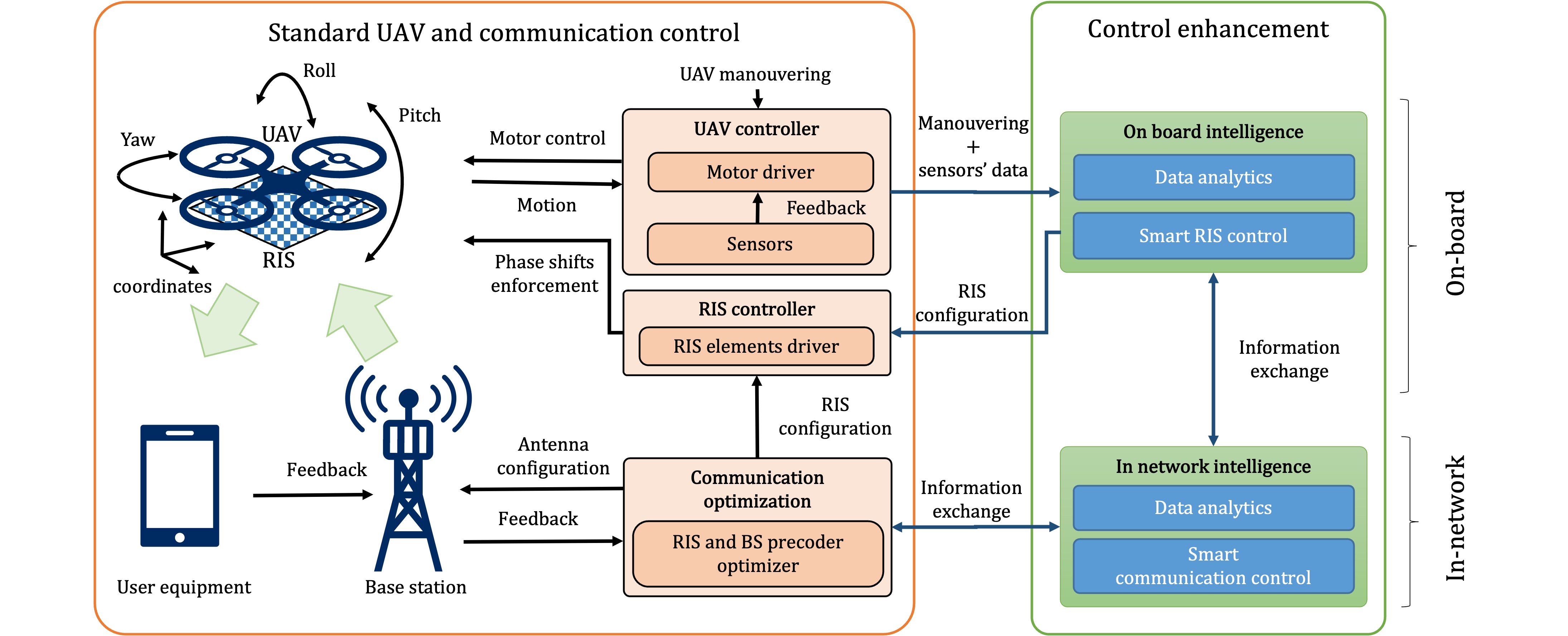}
    \caption{Standard communication and control framework for \acrshort{ris}-aided \acrshort{uav} scenario, with additional control enhancement modules and interfaces enabling advanced \gls{a2g} control.}
    \label{fig:scenario}
\end{figure*}

As depicted on the left-hand side of Fig.~\ref{fig:scenario}, \gls{soa} control architectures for \gls{a2g} networks in the case of \glspl{uav} are typically divided into two separate layers: ground control station and aerial platform control. The former is physically located in the terrestrial network and it consists of a processing unit that, given a set of policies and \gls{qos} requirements, jointly optimizes the \gls{uav} trajectory and both the active and passive beamforming at the \gls{bs} and at the \gls{ris}, respectively. Moreover as described above, it deals with acquiring \gls{csi} and extracting the associated relevant channel parameters. Whereas, the aerial platform control is given by the on-board \gls{uav} controller and the \gls{ris} controller. While the former is dedicated to the maneuvering of the \gls{uav}, the latter triggers \gls{ris} settings (i.e., predefined phase shifts).

A widely adopted assumption in \gls{soa} control architectures is to consider the \gls{uav} in a predefined location in space with negligible orientation and position variations during the communication phase, while its position is updated only within the displacement phase.
However, such assumption does not hold in practical scenarios, wherein the \gls{uav} maneuvering and several atmospheric phenomena can change the position and the orientation of the \gls{uav} even during the communication operations, leading \gls{soa} solutions to be potentially inefficient---or even unfeasible---to operate in realistic conditions. Indeed, as the \gls{uav} is hovering at a certain altitude, its motion is influenced by a \textit{deterministic} component, which is due to the \textit{intentional} maneuvering of the \gls{uav}, i.e., following a predefined trajectory, and a \textit{random} component, due to \textit{unpredictable} factors such as atmospheric conditions including wind, rain, and humidity, imprecise maneuvering, non-ideal \gls{uav} instrumentation, etc.
Such movements result in translations and rotations of the surface of the on-board \gls{ris}, which in turn lead to misalignment of the transmit and reflected beams. This effect is further exacerbated by the highly directive nature of \gls{mmw} beamforming at the \gls{ris} and can ultimately result in loss of connectivity at the user-side~\cite{Mur21}.

To get the most from \gls{a2g} networks in a practical scenario, the mitigation of \gls{uav} mobility effects on the \gls{qos} is a key point. It is thus essential to design enhanced control architectures enabling a transmission optimization tightly coupled with the \gls{uav} mobility pattern. Indeed, \glspl{uav} are equipped with different sensors, such as gyroscope, compass, \gls{gps}, etc., that provide the \gls{uav} controller with motion feedback enabling hoovering control and stabilization. However, due to the separation of the \gls{uav} and \gls{ris} controllers, information on \gls{uav} movements is only partially considered during the system optimization phase, i.e., only the nominal position of the \gls{uav} is considered to perform \gls{csi} acquisition and joint trajectory and beamforming optimization, while information such as real-time maneuvering instructions and \gls{uav} sensors' output is typically neglected. Fig.~\ref{fig:scenario} shows the framework that we envision to enable interaction between the \gls{uav} and the control architecture. In particular, on the right-hand side we consider two distinct \gls{ai}-based modules dubbed as \textit{on-board intelligence} and \textit{in-network intelligence}, which are located on the \gls{uav}, and at the network-side, respectively.
The on-board module interacts both with the \gls{uav} and the \gls{ris} controllers. It plays the fundamental role of collecting mobility information from the \gls{uav} and integrating it into the communication optimization process.
Interestingly, the in-network intelligence module communicates both with the on-board module and with the standard communication optimization module. 
Our proposed novel framework is thus capable of blending together the various conventionally separated system entities and effectively utilizing all the available precious information to suitably optimize both \gls{ris} and \gls{bs} parameters such as, e.g., beamforming configurations and transmit power at the \gls{bs}.

\section{Key-design aspects}
\label{sec:degrees_of_freedom}

\begin{figure}
    \centering
    \includegraphics[width=0.9\linewidth]{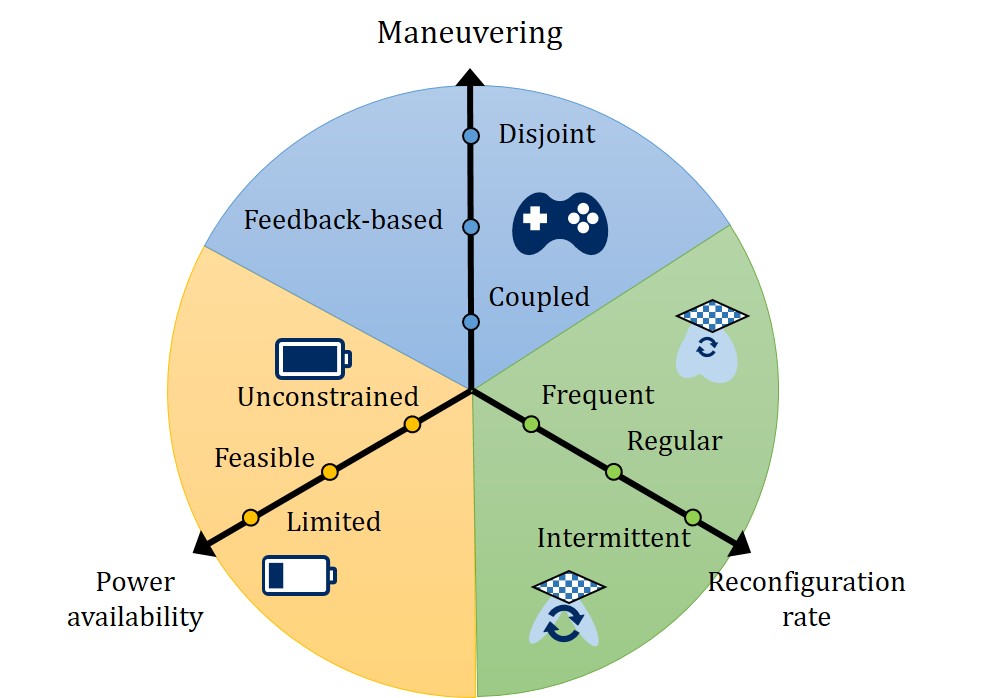}
    \caption{Design choices for the \gls{a2g} enhanced control architecture.}
    \label{fig:degrees_of_freedom}
\end{figure}

To describe our envisioned enhanced control architecture, we first identify the key aspects that significantly affect its design. As depicted in Fig.~\ref{fig:degrees_of_freedom}, an \gls{a2g} network can explore three main directions: \textit{i)} the total available power budget, \textit{ii)} the chosen reconfiguration rate, i.e., the rate at which the key system parameters---such as the \gls{uav} position, the beamforming vectors at the \gls{bs} and the \gls{ris}, the \gls{csi}, etc---need to be updated, and \textit{iii)} the maneuvering control, i.e., identifying which entity in the system is controlling the \gls{uav} and how this interacts with the rest of the network. In the following, we describe the impact of each one of the aforementioned orthogonal directions, which are tightly coupled with the chosen hardware to be employed at both the \gls{uav} and the \gls{ris}, and the desired application scenario, with its associated physical propagation conditions and requirements in terms of \gls{qos}.

{\bf Power availability.} The specific \gls{uav} hardware directly determines the available power budget on board and the percentage used for flying or hovering. In addition, the \gls{uav} may be proprietary of the network operator or might be outsourced from a third-party to lower capital expenditures. In the first case, all the available power is dedicated to optimizing the underlying \gls{a2g} network, whereas in the second case the \gls{uav} movements are independently controlled by an external entity. Indeed, nowadays \glspl{uav} are used for a plethora of different applications such as package delivery or surveillance and it is thus possible to equip them with \glspl{ris} that can be opportunistically used by the network. On the other hand, the size and weight of the \gls{ris} affect the total power consumption, while its number of antenna elements defines the degree of computational power and overhead at the \gls{ris} controller, and at the network-side, e.g., to determine the passive beamforming configuration. In addition, the chosen application scenario plays a fundamental role in the total power availability. In particular, the number of users and their mobility behaviors, and the propagation conditions such as the presence of obstacles or unwanted meteorological phenomena are key aspects to be carefully considered. Moreover, the \gls{uav} needs to be re-charged in appropriate charging stations, whose location constraints the total power availability in the environment. Such aspects have a direct impact on the joint beamforming and trajectory optimization of the \gls{ris} and the \gls{uav}, respectively, and thus imply a given computational power. The latter is also directly influenced by the required \gls{qos}: in disaster situations the focus is on sending emergency time-critical signals that are characterized by low latency and low rate, thus giving rise to low computational power, whereas less critical applications such as data streaming may accept a larger latency but require a higher rate, which is typically associated with an increased computational power. In this regard, we identify three categorizations namely \textit{unconstrained, feasible,} and \textit{limited} power availability.

{\bf Reconfiguration rate.} The choice of reconfiguration rates enables different \gls{qos} guarantees and overall achievable communication rate. We determine three degrees of increasing reconfiguration rate dubbed as \textit{intermittent}, \textit{regular}, and \textit{frequent}. For low reconfiguration rates, the system can devote most of the available time for sending/receiving data to/from the users. However, the available information from the \gls{uav} sensors, user feedback, and the \gls{csi} acquisition is only sporadically updated. As a result, the performance might be acceptable only for low mobility scenarios or when the atmospheric conditions are ideal. Whereas for frequent reconfiguration rate the available information is continuously updated and as a result, the communication quality can be generally maximized even under high user mobility and strong meteorological perturbations. However, this implies an increased overhead that negatively affects the overall data rate. An intermediate solution is given by regular reconfiguration rates, which is the case of robust optimization algorithms, i.e., schemes that employ statistical channel and perturbation information rather than costly instantaneous \gls{csi}.

{\bf Maneuvering.} The \gls{uav} maneuvering is either handled by the network itself, i.e., it is \textit{coupled} with the optimization of the system, or by an operator like in disaster situations, in which the information about the \gls{uav} position is \textit{feedback based} from the control device. In such scenarios, the network can access useful information from the \gls{uav} sensors and can accurately track the \gls{csi} as the \gls{uav} moves. In the second case, the \gls{uav} maneuvering might be under the control of an external entity and is thus \textit{disjoint} from the rest of the system optimization. Hence, the \gls{csi} and the \gls{uav} position must be regularly estimated in order to keep an acceptable level of performance. 


\section{RIS-aided A2G Case Studies for 6G}
\label{sect:case-study}

\begin{figure*}
    \centering
    \includegraphics[width=1\linewidth]{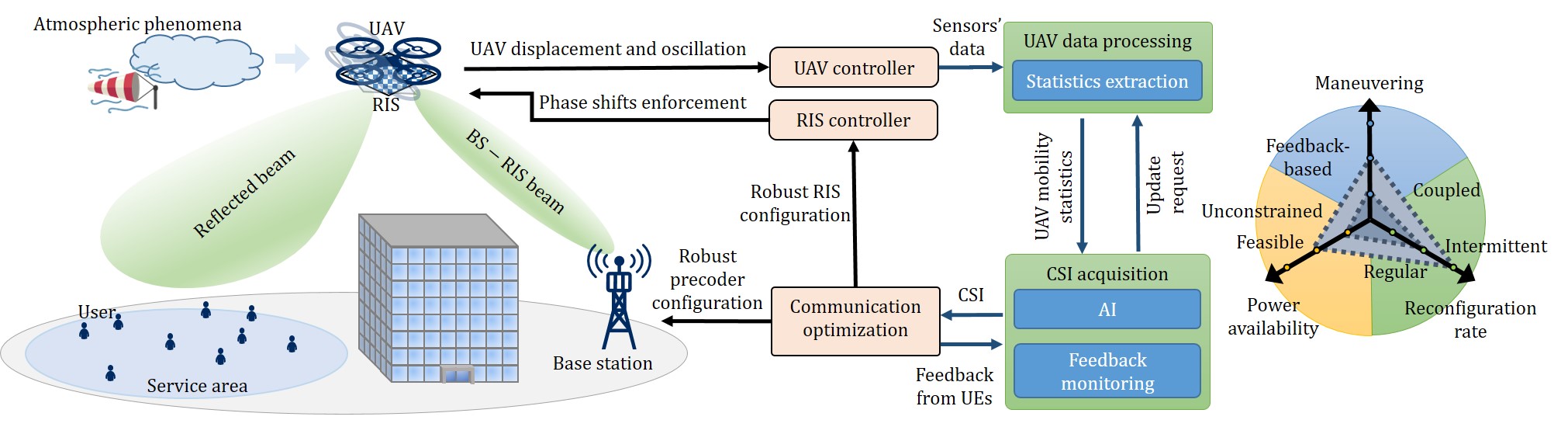}
    \caption{Practical scenario with static \gls{uav} subject to undesired displacement and oscillation due to atmospheric phenomena. The output of the \gls{uav} sensors are processed and used to obtain a \gls{csi} representation accounting for the \gls{uav} mobility statistics. The \gls{ris} configuration is optimized accordingly to preserve the communication quality.}
    \label{fig:fixed_uav_scenario}
\end{figure*}

Hereafter, we describe a practical implementation of our proposed enhanced control architecture considering two relevant case studies when \gls{ris} technology is in place.

{\bf Case study: Static UAV.} The relevant scenario of interest is depicted in Fig.~\ref{fig:fixed_uav_scenario}. Assume that due to low mobility of ground users a \gls{uav} is in a fixed location in space, which is only seldomly updated. However, due to adverse atmospheric conditions, the \gls{uav} is subject to unwanted perturbations, which result in undesired roll, yaw, and pitch of the surface of the on-board \gls{ris}. \gls{uav} movement counteractions are automatically taken but still orientation oscillations or location perturbation may result in an instantaneously \gls{ris} misconfiguration, leading to misalignment of the reflected beams and degraded overall achievable rate. Remarkably, it has been recently shown that it is possible to guarantee an acceptable level of performance in the target area of influence by suitably adjusting the beamforming configuration as a function of the second-order statistics of such perturbations~\cite{Mur21}.
Such adaptation of the system configuration is enabled by our proposed enhanced control architecture. Indeed, the measurements of the instantaneous roll, yaw and pitch are collected by the \gls{uav} controller and sent to the \gls{uav} data processing module that extracts or predicts (e.g., using \gls{ai}) the relevant statistics. This information is then used upon request to update the current \gls{csi} and optimize the communication to/from the \gls{uav}. In particular, the beamforming configuration can be optimized on the basis of the current perturbation statistics via:  i) conventional mathematical tools such as \gls{sdr}, ii) by training a \gls{ml} model that learns how to adapt the beamforming configuration to the varying atmospheric conditions, or iii) by designing an \textit{online} \gls{ai} learning algorithm. The choice of optimization method determines the computational power expenditure, which has a direct impact on the total power availability. As a result, our proposed enhanced control architecture can be designed to have a \textit{feasible} or \textit{unconstrained} power availability.

As the \gls{uav} is in a fixed position, the \gls{ris} and \gls{bs} precoders can be updated only when needed, i.e., when the perturbation statistics evolve due to a change of atmospheric conditions. Our proposed enhanced control architecture provides two ways to deal with such a scenario, which are characterized by an \textit{intermittent} and \textit{regular} reconfiguration rate, respectively. A first approach is to exploit feedback monitoring: users in the service area can periodically send updates to the network with the current perceived \gls{qos}. The in-network intelligence devoted to the \gls{csi} acquisition issues a statistics update request to the \gls{uav} data processing module if relevant changes in the \gls{qos} are detected. 

A second approach is based on proactively sending updates on the \gls{uav} perturbation statistics when a relevant change is detected from the \gls{uav} data processing module to the \gls{csi} acquisition module. The latter then decides whether the relevant system parameters, such as the beamforming configurations, should be updated. This approach leads to higher power consumption at the \gls{uav} due to continuous monitoring and increased communication overhead. On the other hand, it allows to quickly react to the varying environmental conditions and minimize network downtime.

Lastly, note that both aforementioned approaches can be realized under \textit{coupled} or \textit{feedback-based} maneuvering. Indeed, in such cases, the \gls{uav} is controlled by the network and can thus retrieve useful information from the \gls{uav} controller. Whereas, a \textit{disjoint} implementation would make it unfeasible to extract statistical information on the unwanted perturbations of the \gls{uav} position.

\begin{figure*}
    \centering
    \includegraphics[width=1\linewidth]{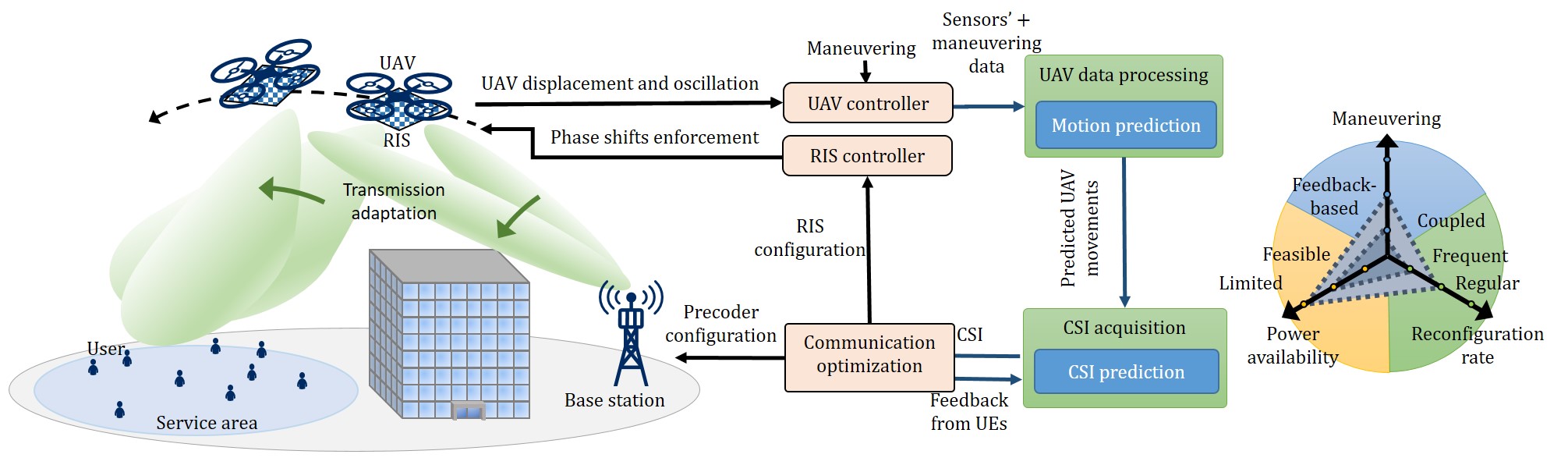}
    \caption{Practical scenario with \gls{uav} subject to intentional maneuvering operations with corresponding displacement and orientation changes. The maneuvering operation and the output of the \gls{uav} sensors is processed and used to obtain a prediction of the \gls{uav} position and orientation in time. The \gls{ris} configuration is optimized according to the \gls{uav} motion to enable transmission adaptation.}
    \label{fig:moving_uav_scenario}
\end{figure*}

{\bf Case study: Nomadic UAV.} We consider the practical case where a \gls{uav} is subject to desired movements in space, i.e., following a predetermined trajectory, as depicted in Fig.~\ref{fig:moving_uav_scenario}. For simplicity, we neglect user mobility in order to focus on providing coverage enhancement within a given target service area. 
The drone movements lead to a continuous position and orientation change of the on-board \gls{ris}. Such effect, if not properly addressed, could lead to severe beam misalignment, and potentially to complete service disruption. Therefore, to maintain a stable connection, a continuous adaptation of both the \gls{ris} configuration and the \gls{bs} precoder is required.

Referring to the degrees of freedom highlighted in Section~\ref{sec:degrees_of_freedom}, we consider the \gls{uav} maneuvering to be either \emph{coupled}, i.e., the \gls{uav} trajectory is imposed by decisions taken at the network-side and therefore jointly optimized with the beamforming strategy, or \emph{feedback-based}, i.e., the \gls{uav} is controlled by an external operator (e.g., a member of first responder teams) and therefore, its movements are not perfectly known to the network. 


In the case of \emph{coupled} maneuvering, the communication optimization module can compute in advance the \gls{ris} and the \gls{bs} beamforming configuration according to the \gls{uav} trajectory evolution. Meanwhile, thanks to the \gls{uav} sensors data, the on-board data processing module can track the effective trajectory evolution, compare it against the desired one, and feed back information in case of divergence (e.g., non-idealities of the \gls{uav} controller, wind, etc.). Thus enabling suitable adjustment of both the beamforming and the maneuvering, and improving the overall reliability and robustness of the system. Depending on the power availability and variety of sensors equipping the \gls{uav}, the on-board data processing module can be further exploited. For example, visual information from cameras could be used to perform object detection and reveal potential obstruction (e.g., buildings, trees, or debris) and adapt the \gls{uav} trajectory accordingly.

Whereas in the case of \emph{feedback-based} maneuvering, the \gls{uav} movements are known at the network-side only a-posteriori. Therefore, to prevent the communication optimization from lagging behind the \gls{uav} movements, a trajectory prediction strategy could be applied to the on-board data processing module. Such prediction could take advantage of both the sensor data and the maneuvering feedback, while the trajectory forecast could be sent to the network-side to compute the optimal beamforming configuration in advance, so as to enable a more accurate transmission adaptation to the \gls{uav} movement.

The need for continuous communication adaptation, specific of the nomadic \gls{uav} scenario, results in a control architecture design that has to sustain a relatively high reconfiguration rate, i.e., \emph{regular} or \emph{frequent}.
This implies a potentially high control overhead to frequently transmit the \gls{ris} configuration to the \gls{ris} controller. Nonetheless, smart overhead reduction strategies can be implemented thanks to the proposed enhanced control architecture. Indeed, the reconfiguration can be avoided as long as the \gls{qos} is within the desired level, despite the movement of the \gls{uav}. Therefore, a rate adaptation strategy could be considered to trigger the reconfiguration only when needed, e.g. based on the users' feedback monitoring, thus minimizing the communication overhead.
Moreover, the \gls{ris} configuration typically exhibits regular and periodic patterns over the surface elements. This peculiarity can be exploited by advanced encoding and decoding techniques, e.g., autoencoders, that could be easily implemented in-network (encoding) and on-board (decoding) to further reduce the overhead. Such an additional feature would stress the computational load of the on-board intelligence module and in turn reflect on the power availability requirements. As a result, this scenario is characterized by \textit{limited} or \textit{feasible} power availability.

\begin{figure}
    \centering
    \includegraphics[width=1\linewidth]{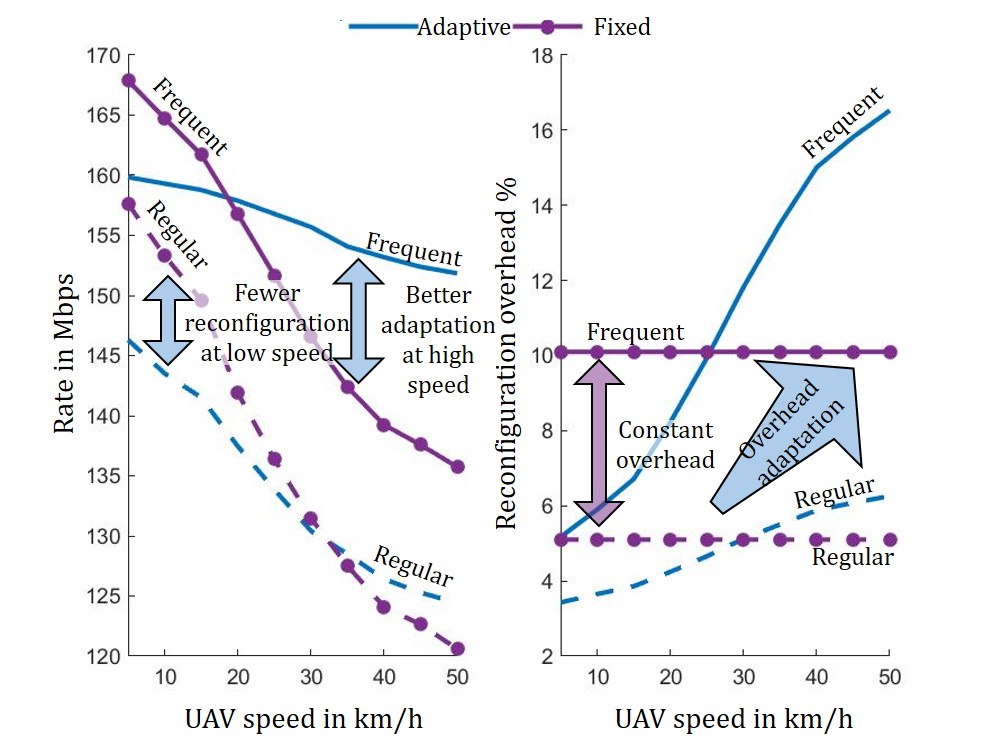}
    \caption{Comparison of the achievable rate and the control overhead in a nomadic \gls{uav} scenario obtained with standard periodic \gls{ris} reconfiguration and smart reconfiguration adaptation enabled by the enhanced control architecture versus the \gls{uav} speed.}
    \label{fig:feedback_vs_performance}
\end{figure}

\section{Discussion}

The performance of our proposed enhanced control architecture in the application scenario described in Section~\ref{sect:case-study} are hereafter summarized, for the case of \textit{feedback-based} maneuvering, i.e., in the presence of an external \gls{uav} operator. We consider an \gls{a2g} network including one single-antenna \gls{bs}, one \gls{uav} equipped with a $100$ element squared \gls{ris} and a target single-antenna user located at a distance of $70$ meters from the \gls{bs}. We assume the operator moves the \gls{uav} following a trajectory encircling the user with a radius of $25$ m, altitude of $20$ m, and with a given variable speed. The working frequency is set to $30$ GHz, while the transmit power at the \gls{bs} is fixed to $24$ dBm and the noise spectral density is assumed to be $-80$ dBm. 

In Fig.~\ref{fig:feedback_vs_performance}, we compare two different schemes, namely our proposed enhanced control architecture dubbed as \textit{Adaptive} and the standard \gls{soa} control framework namely \textit{Fixed}. We vary the \gls{uav} speed from $5$ to $50$ km/h and consider two different choices of reconfiguration rate, denoted as \textit{Frequent} in solid lines and \textit{Regular} in dashed lines. Thanks to our proposed control architecture, the in-network intelligent modules receive feedback from the \gls{uav} maneuvering, the \gls{uav} sensors and the user-perceived \gls{qos} and use it to adapt the system reconfiguration rate to the current \gls{uav} mobility rate and effective trajectory. In particular, as shown on the right-hand side, our proposed scheme adapts the \gls{csi} and \gls{ris} beamforming strategy more frequently for an increasing \gls{uav} speed. On the one hand this generates higher overhead, which is shown as a percentage of the total available transmission time. On the other hand, the data rate is kept high and quasi-constant for the case of \textit{frequent} reconfiguration rate thanks to the up-to-date \gls{csi}, which in turn leads to high \gls{snr}.

In contrast to this, the standard architecture does not have access to the aforementioned \gls{uav} and \gls{qos} status information and thus, uses a pre-determined refresh rate to perform \gls{csi} acquisition and \gls{ris} beamforming optimization, as is typically the case for directional communications such as \gls{mmw}. In this case, the communication overhead is constant versus the \gls{uav} speed, while the data rate monotonically decreases. However, at low \gls{uav} speeds, the standard scheme obtains higher data rates at the cost of an increased overhead as compared to the proposed enhanced scheme.








\section{Conclusions}
Aerial communications are opening a new research direction to enable a \gls{3d} mobile networking paradigm expected to be effective in the 6G landscape. {However,} a number of daunting challenges need to be addressed before the dream of cost-effective \emph{flying mobile stations} solutions can be reached{, in particular with respect to the design of the control architecture}.

In this paper, we analyzed the novel concept of \gls{ris}-aided aerial communications and shed the light on some of its potential use cases, optimization aspects, and challenges. Specifically, we argued that a carefully designed enhanced control architecture is essential in order to take full advantage of the \gls{ris} \gls{3d} passive beamforming capabilities and the flexibility of flying devices. In contrast to existing \gls{soa} frameworks, our envisioned enhanced control architecture is able \emph{to bridge together several conventionally isolated entities and exploit useful information generated by the \gls{uav} sensors} (in addition to the user feedback on the perceived \gls{qos}) in order to adapt the system configuration to varying environmental conditions.

We analyzed two practical scenarios of interest wherein the underlying \gls{a2g} network benefits from our proposed architecture and quantified its performance in a relevant case study. Our results show that $i$) our proposed \emph{Adaptive} scheme is able to adapt the \gls{csi} and \gls{ris} beamforming strategy as required according to varying \gls{uav} speeds, $ii$) the \emph{adaptability} comes at a non-negligible reconfiguration overhead cost of about $5$ to $15$\% and $iii$) the proposed \emph{adaptive} solution successfully manages to keep the data rate at low degradation percentages ($<10\%$) for the UAV speed range considered ($5$ to $50$ km/h).




\bibliographystyle{IEEEtran}
\bibliography{references}

\section*{Biographies}
\vskip -2\baselineskip plus -1fil
\begin{IEEEbiographynophoto}{Francesco Devoti} (M’20) received the B.S., and M.S. degrees in Telecommunication Engineering, and the Ph.D. degree in Information Technology from the Politecnico di Milano, in 2013, 2016, and 2020 respectively. He is currently a research scientist at NEC Laboratories Europe. His research interests include millimeter-wave technologies in 5G networks.
\end{IEEEbiographynophoto}
\vskip -2\baselineskip plus -1fil

\begin{IEEEbiographynophoto}{Placido Mursia} (S'18--M'21) received the B.Sc. and M.Sc. (with honors) degrees in Telecommunication Engineering from Politecnico of Turin in 2015 and 2018, respectively. He obtained his Ph.D from Sorbonne Université of Paris, at the Communication Systems department of EURECOM in 2021. He is currently a research scientist in the 5GN group at NEC Laboratories Europe. His research interests lie in convex optimization, signal processing and wireless communication.
\end{IEEEbiographynophoto}
\vskip -2\baselineskip plus -1fil

\begin{IEEEbiographynophoto}{Vincenzo Sciancalepore} (S'11--M'15--SM'19) received his M.Sc. degree in Telecommunications Engineering and Telematics Engineering in 2011 and 2012, respectively, whereas in 2015, he received a double Ph.D. degree. Currently, he is a Principal Researcher at NEC Laboratories Europe, focusing his activity on reconfigurable intelligent surfaces. He is an Editor of the IEEE Transactions on Wireless Communications.
\end{IEEEbiographynophoto}
\vskip -2\baselineskip plus -1fil

\begin{IEEEbiographynophoto}{Xavier Costa-Perez} (M'06--SM'18) is Research Professor at ICREA, Scientific Director at the i2Cat R\&D Center and Head of 6G Networks R\&D at NEC Laboratories Europe. 
Xavier served on the Program Committee of several conferences (including IEEE ICC and INFOCOM), published at top research venues and holds several patents. 
He received his M.Sc. and Ph.D. degrees in Telecommunications from the Polytechnic University of Catalonia (UPC) in Barcelona.
\end{IEEEbiographynophoto}
\end{document}